\documentstyle[12pt,a4]{article}
\input epsf

\global\arraycolsep=2pt

\begin{document}

\begin{titlepage}

\begin{flushright}
CERN-TH/96-282\\
hep-ph/9610471
\end{flushright}

\vspace{0.5cm}
\begin{center}
\Large\bf Exploring the Invisible Renormalon:\\
Renormalization of the Heavy-Quark Kinetic Energy
\end{center}

\vspace{1.5cm}
\begin{center}
Matthias Neubert\\
{\sl Theory Division, CERN, CH-1211 Geneva 23, Switzerland}
\end{center}

\vspace{1.0cm}
\begin{abstract}
Using the virial theorem of the heavy-quark effective theory, we show
that the mixing of the operator for the heavy-quark kinetic energy
with the identity operator is forbidden at the one-loop order by
Lorentz invariance. This explains why such a mixing was not observed
in several one-loop calculations using regularization schemes with a
Lorentz-invariant UV regulator, and why no UV renormalon singularity
was found in the matrix elements of the kinetic operator in the
bubble approximation (the ``invisible renormalon''). On the other
hand, we show that the mixing is not protected in general by any
symmetry, and it indeed occurs at the two-loop order. This implies
that the parameter $\lambda_1^H$ of the heavy-quark effective theory
is not directly a physical quantity, but requires a non-perturbative
subtraction.
\end{abstract}

\vspace{1.0cm}
\centerline{(Submitted to Physics Letters B)}

\vspace{2.5cm}
\noindent
CERN-TH/96-282\\
October 1996

\end{titlepage}

\section{Introduction}

The physics of hadrons containing a heavy quark simplifies greatly in
the limit where the heavy-quark mass $m_Q$ is taken to infinity. Then
new symmetries of the strong interactions arise, which relate the
long-distance properties of many observables to a small number of
hadronic matrix elements \cite{review}. A systematic expansion around
the heavy-quark limit is provided by the heavy-quark effective theory
(HQET), in which a heavy quark inside a hadron moving with
four-velocity $v$ is described by a velocity-dependent field $h_v(x)$
subject to the constraint $\rlap/v\,h_v=h_v$. The field $h_v(x)$ is
related to the original heavy-quark field $Q(x)$ by a phase
redefinition, which removes the large ``mechanical'' part $m_Q v$ of
the heavy-quark momentum arising from the motion of the heavy hadron.
Thus, the effective field carries the ``residual momentum''
$k=p_Q-m_Q v$, which characterizes the interactions of the heavy
quark with gluons. The effective Lagrangian of the HQET is
\cite{EiHi}--\cite{FGL}
\begin{equation}
   {\cal L}_{\rm eff} = \bar h_v\,i v\!\cdot\!D\,h_v
   + \frac{1}{2 m_Q}\,\bar h_v\,(i D_\perp)^2 h_v
   + \frac{C_{\rm mag}\,g_s}{4 m_Q}\,
   \bar h_v\,\sigma_{\mu\nu} G^{\mu\nu} h_v + O(1/m_Q^2) \,,
\label{Leff}
\end{equation}
where $D^\mu=\partial^\mu-i g_s A^\mu$ is the gauge-covariant
derivative, and $D_\perp^\mu=D^\mu-(v\cdot D)\,v^\mu$ contains its
components orthogonal to the velocity. The origin of the operators
arising at order $1/m_Q$ is most transparent in the rest frame of the
heavy hadron: the first operator corresponds to the kinetic energy
resulting from the motion of the heavy quark inside the hadron (in
the rest frame, $(i D_\perp)^2$ is the operator for $-{\bf k}^2$),
whereas the second operator describes the magnetic interaction of the
heavy-quark spin with the gluon field. The Wilson coefficient $C_{\rm
mag}$ results from short-distance effects and depends logarithmically
on the scale at which the chromo-magnetic operator is renormalized.
As a consequence of the so-called reparametrization invariance of the
HQET (an invariance under infinitesimal changes of the velocity), the
kinetic operator is not multiplicatively renormalized
\cite{LuMa,Chen}.

The matrix elements of the two dimension-five operators in
(\ref{Leff}) play a most significant role in many applications of the
HQET. They appear, for instance, in the spectroscopy of heavy hadrons
and in the description of inclusive weak decays \cite{review}. For
the hadronic matrix elements of the kinetic operator, in particular,
one defines hadronic parameters $\lambda_1^H$ by\footnote{Another
common notation is to define $\mu_{\pi,H}^2=-\lambda_1^H$.}
\cite{FaNe}
\begin{equation}
   \langle H(v)|\,\bar h_v\,(i D_\perp)^2 h_v\,|H(v)\rangle
   = \lambda_1^H \,,
\end{equation}
where we use a mass-independent normalization of states such that
$\langle H(v)|\,\bar h_{v} h_v\,|H(v)\rangle=1$. Spectroscopic
relations may be used to extract the difference in the values of
$\lambda_1^H$ for two hadron states. For the ground-state heavy
mesons and baryons, for instance, we obtain
\begin{equation}
   \frac{m_{\Lambda_b}-m_{\Lambda_c}}{\overline{m}_B-\overline{m}_D}
   = 1 + \frac{\lambda_1^{\rm bar} - \lambda_1^{\rm mes}}
    {2 \overline{m}_D \overline{m}_B} + O(1/m_H^3) \,,
\label{lamdif}
\end{equation}
where $O(1/m_H^3)$ means terms suppressed by three powers of the
large hadron masses, and $\overline{m}_B=\frac{1}{4}\,(m_B + 3
m_{B^*})$ and $\overline{m}_D=\frac{1}{4}\,(m_D + 3 m_{D^*})$ denote
the spin-averaged meson masses, defined such that they do not receive
a contribution from the chromo-magnetic interaction. The value of a
single parameter $\lambda_1^H$ cannot be determined from
spectroscopy, since it always appears in combination with the
heavy-quark mass, which by itself is not a physical parameter. This
observation poses the questions about the status of the heavy-quark
kinetic energy as a physical parameter.

In this letter, we study the mixing of the kinetic operator $\bar
h_v\,(iD_\perp)^2 h_v$ with the ``identity operator'' $\bar h_v h_v$
under ultraviolet (UV) renormalization. By naive dimensional
analysis, such a mixing is expected to occur in regularization
schemes with a dimensionful cutoff parameter $\lambda$, since the two
operators have the same quantum numbers but different dimension. This
would have important implications for phenomenology, as it leads to
an additive contribution to the parameter $\lambda_1^H$ of the form
$\lambda^2\,C[\alpha_s(\lambda)]$, which must be subtracted in order
to define a renormalized (``physical'') parameter that is independent
of the UV regulator. The coefficient $C[\alpha_s(\lambda)]$ can be
calculated order by order in an expansion in the small coupling
constant $\alpha_s(\lambda)$, and it appears at first sight that the
quadratically divergent term could be subtracted using perturbation
theory. This impression is erroneous, however, because $C$ may
contain non-perturbative contributions of the form
$\exp[-8\pi/\beta_0\alpha_s(\lambda)]=(\Lambda_{\rm QCD}/\lambda)^2$,
which cannot be controlled in perturbation theory \cite{MMS}. Such
terms can contribute an amount of order $\Lambda_{\rm QCD}^2$ to the
parameter $\lambda_1^H$, which is of the same order as the
renormalized parameter itself. Hence, if the kinetic operator mixes
with the identity operator, it is necessary that the quadratically
divergent contribution to $\lambda_1^H$ be subtracted in a
non-perturbative way, and hence the heavy-quark kinetic energy by
itself is not directly a physical quantity. On the other hand, since
the quadratic divergence is an UV effect and therefore insensitive to
the nature of the external states, it follows that the difference in
the values of $\lambda_1^H$ for two hadrons is a physical quantity,
and relations such as (\ref{lamdif}) are meaningful.

The issue of power divergences of matrix elements is intimately
related to that of UV renormalons \cite{CTS}. The purpose of the
heavy-quark expansion is to disentangle the short-distance physics
characterized by the large mass scale $m_Q$ from the long-distance
physics characterized by the typical scale of the momenta exchanged
between the heavy quark and light degrees of freedom. This is
achieved by introducing a factorization scale $\lambda$ such that
$\Lambda_{\rm QCD}\ll\lambda\ll m_Q$. Contributions from momenta
above $\lambda$ are controllable in perturbation theory and
attributed to Wilson coefficients, whereas contributions from momenta
below $\lambda$ are contained in the matrix elements of the operators
in the HQET. If this program is performed with a ``hard''
factorization scale, these matrix elements diverge, for dimensional
reasons, as powers of the UV cutoff. For practical reasons, however,
one usually calculates the Wilson coefficients using dimensional
regularization. In this case power divergences do not appear, and the
factorization scale $\lambda$ is replaced by a renormalization scale
$\mu$. It is then unavoidable that the Wilson coefficients receive
contributions from momenta below $\mu$ (so-called infrared (IR)
renormalons), and the operator matrix elements receive contributions
from momenta above $\mu$ (so-called UV renormalons). These
contributions lead to a factorial growth of the coefficients in the
perturbative expansion of the Wilson coefficients and matrix
elements. The corresponding perturbation series are divergent and not
Borel summable; they must be truncated close to the minimal term.
Therefore, the presence of renormalons leads to intrinsic ambiguities
in the definition of the Wilson coefficients and matrix elements,
which only cancel when all short- and long-distance contributions are
combined in the heavy-quark expansion \cite{Davi}--\cite{LMS}. It is
generally believed that from the degree of divergence of the matrix
elements of an operator in the HQET one can deduce the position of
the corresponding UV renormalon singularities in the Borel transform
of these matrix elements with respect to the inverse coupling
constant; a power divergence proportional to $\lambda^n$ corresponds
to a singularity at the position $u=n/2$ (in a certain normalization
\cite{Chris}) on the positive real axis in the Borel plane. For the
case of the self-energy of a heavy quark, for instance, the
correspondence between a linear UV divergence and a renormalon pole
at $u=\frac 12$ has been established in \cite{BeBr,Bigi}.

The question whether there is a mixing of the kinetic energy with the
identity operator, and whether there exists a corresponding
renormalon singularity, has been addressed previously by several
authors, with seemingly controversial conclusions. At the one-loop
order, such a mixing has indeed been observed when the HQET
regularized on a space--time lattice \cite{MMS}. Likewise, a
``physical'' definition of a parameter $\lambda_1^H(\lambda)$ has
been suggested, which absorbs certain $O(\alpha_s)$ corrections
appearing in the zero-recoil sum rules for heavy-quark transitions
\cite{Bigsr}. This definition is such that
$\mbox{d}\lambda_1^H(\lambda)/\mbox{d}\lambda^2\propto
\alpha_s(\lambda)$, indicating again a one-loop mixing of the kinetic
energy with the identity operator. On the other hand, this mixing has
not been observed at the one-loop order in two Lorentz-invariant
cutoff regularization schemes \cite{MNS}, which use a Pauli--Villars
regulator or a cutoff on the virtuality of the gluon in one-loop
Feynman diagrams \cite{flow}. This observation appeared as a puzzle,
because there seemed to be no obvious reason why the mixing should
not occur at the one-loop order, and indeed individual diagrams were
quadratically divergent. Yet, the vanishing of the operator mixing at
the one-loop order has found its counterpart in the analysis of UV
renormalons. As mentioned above, a quadratic divergence would
correspond to a renormalon singularity at $u=1$ in the Borel plane.
Even though we are far from being able to derive an exact expression
for the Borel transform of a perturbation series, explicit results
can be obtained in the so-called ``bubble approximation'', in which
all terms corresponding to an arbitrary number of self-energy
insertions on a gluon propagator are kept. Surprisingly, it was found
that in this approximation the renormalon singularity at $u=1$ in the
matrix elements of the kinetic operator is absent \cite{MNS}. The
absence of the related IR renormalon in the pole mass of a heavy
quark was noted previously in \cite{BeBr}. This ``missing''
singularity has been called the ``invisible renormalon''. Three
possible explanations for this puzzle have been suggested:
\begin{enumerate}
\item
The vanishing of the operator mixing at the one-loop order is
accidental and happens only in some peculiar regularization schemes.
However, this does not explain why the renormalon singularity at
$u=1$ is ``invisible'' in the bubble approximation.
\item
The vanishing of the mixing is an artefact of one-loop perturbation
theory. The mixing appears in higher orders, and the corresponding
renormalon singularity appears when one goes beyond the bubble
approximation.
\item
There is a symmetry which prevents the mixing between the kinetic
energy and the identity operator, but this symmetry is broken by the
lattice regularization. Candidates of such symmetries are Lorentz (or
rotational) invariance and the reparametrization invariance of the
HQET \cite{LuMa,Chen}. In fact, it has been conjectured that Lorentz
invariance may protect the matrix elements of the kinetic energy
against an UV renormalon singularity at $u=1$ \cite{BBZ}. However,
later it was argued that there is no obvious connection between
Lorentz or reparametrization invariance and the absence of a
quadratic divergence (or the corresponding UV renormalon) in the
matrix elements of the kinetic energy \cite{MNS}.
\end{enumerate}

In this paper we shall resolve this puzzle. We will prove that the
mixing between the kinetic energy and the identity operator is
forbidden at the one-loop order in all regularization schemes with a
Lorentz-invariant UV cutoff. For our arguments to hold, it is
essential that the regulator be introduced in a fully
Lorentz-invariant way; in particular, it is not sufficient to choose
a regularization scheme that respects only rotational invariance in
the rest frame of the heavy hadron. The fact that a mixing at the
one-loop order was observed in \cite{MMS} and \cite{Bigsr} is a
consequence of the explicit breaking of Lorentz invariance by the
regularization schemes adopted in these calculations. On the other
hand, in general there is no symmetry that protects the matrix
elements of the kinetic operator from quadratic divergences, and we
show explicitly that a mixing with the identity operator occurs from
the two-loop order on. In the renormalon language, this suggest that
there is a singularity at $u=1$ in the Borel plane, which would
appear when one goes beyond the bubble approximation. Hence, each of
the above-mentioned suggestions is only partly correct, and the
puzzle of the ``invisible renormalon'' finds its explanation in an
``accidental'' space--time symmetry, which is only operative at the
one-loop order.

\section{Operator Mixing}

\begin{figure}
\epsfxsize=10cm
\centerline{\epsffile{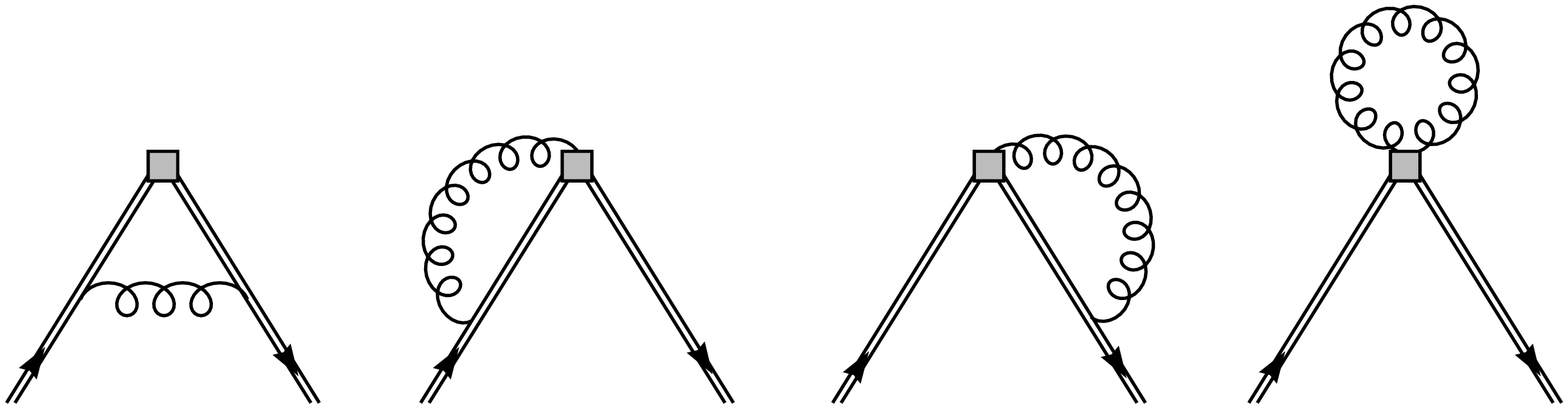}}
\caption{One-loop diagrams that could contribute to the mixing of the
kinetic energy with the identity operator. The kinetic operator is
represented by a grey square; heavy-quark propagators are drawn as
double lines.}
\label{fig:kinop}
\end{figure}

As mentioned above, in regularization schemes with a dimensionful
cutoff $\lambda$, the kinetic operator can mix with the identity
operator under UV renormalization. We define a renormalization
constant $Z_{D^2\to\hat 1}$ by
\begin{equation}
   \bar h_{v}\,(i D_\perp)^2 h_v \Big|_{\rm bare}
   = Z_{D^2\to\hat 1}\,\bar h_v h_v + \dots \,,
\end{equation}
where the subscript ``bare'' indicates the bare operator, whose
matrix elements depend on the UV regulator, and the ellipses
represent contributions from operators of higher dimension. Such a
mixing leads to an additive contribution to the matrix elements of
the kinetic operator which, for dimensional reasons, is quadratically
divergent in the cutoff $\lambda$. It follows that
\begin{equation}
   \frac{\mbox{d}}{\mbox{d}\lambda^2}\,\lambda_1^H(\lambda)
   = \frac{\mbox{d}}{\mbox{d}\lambda^2}\,Z_{D^2\to\hat 1} \,.
\label{lam1ren}
\end{equation}
Our goal is to perform a perturbative calculation of $Z_{D^2\to\hat
1}$ in a regularization scheme where the cutoff preserves Lorentz
invariance. By evaluating the UV divergences of the diagrams shown in
Fig.~\ref{fig:kinop} (supplemented by the wave-function
renormalization of the quark fields), it has been shown in \cite{MNS}
that $Z_{D^2\to\hat 1}$ vanishes at the one-loop order in two such
regularization schemes. However, individual diagrams have a quadratic
divergence, and the reason for the cancellations occurring in the sum
of all diagrams remained unclear.

To extend the direct calculation of $Z_{D^2\to\hat 1}$ to the next
order would involve the evaluation of a large number of two-loop
diagrams. In this letter, we suggest a more efficient way to perform
this calculation, which in addition will provide an explanation for
the vanishing of $Z_{D^2\to\hat 1}$ at the one-loop order. To this
end, we use the virial theorem of the HQET, which establishes a
relation between the kinetic energy and a matrix element of an
operator containing the gluon field-strength tensor between states
moving at different velocities \cite{FaNe,virial}. For the
ground-state pseudoscalar mesons\footnote{The same relation holds for
the ground-state vector mesons, when an averaging over the two
transverse polarization states is implied.} and $\Lambda_Q$ baryons,
the virial theorem states that
\begin{equation}
   \langle H(v')|\,\bar h_{v'} ig_s G^{\mu\nu} h_v\,|H(v)\rangle
   = (v^\mu v'^\nu - v^\nu v'^\mu)\,\left\{
   - \frac{\lambda_1^H}{3} + O(v\cdot v'-1) \right\} \,.
\label{virial}
\end{equation}
This is a non-perturbative relation between hadronic matrix elements,
which is preserved by renormalization \cite{new}. In the rest frame
of one of the hadrons, only the electric field has a non-vanishing
matrix element, and we shall therefore refer to the operator on the
left-hand side as the chromo-electric operator. Note that
(\ref{virial}) is not an operator identity, but rather a relation
between ground-state matrix elements; the corresponding operator
identity would contain operators of higher spin, which have vanishing
ground-state matrix elements. However, (\ref{virial}) is an operator
identity as far as the quadratic UV divergences are concerned.

Similar to the kinetic operator, also the chromo-electric operator
can mix with a dimen\-sion-three operator. Since the Feynman rules of
the HQET do not involve Dirac matrices, there is only one candidate
for such an operator, and we define
\begin{equation}
   \bar h_{v'} ig_s G^{\mu\nu} h_v \Big|_{\rm bare}
   = Z_{E\to\hat 1}(v\cdot v')\,(v^\mu v'^\nu - v^\nu v'^\mu)\,
   \bar h_{v'} h_v + \dots \,.
\label{ZEdef}
\end{equation}
Although there exist other dimension-three operators with the right
quantum numbers, such as $\bar h_{v'}i\sigma^{\mu\nu} h_v$ and $\bar
h_{v'}[(v-v')^\mu\gamma^\nu - (v-v')^\nu\gamma^\mu]\,h_v$, they
cannot be induced by quantum effects. The virial theorem
(\ref{virial}) implies a relation between the renormalization factors
$Z_{E\to\hat 1}$ and $Z_{D^2\to\hat 1}$ in the limit of equal
velocities $(v\cdot v'=1$), which reads
\begin{equation}
   Z_{D^2\to\hat 1} = - 3 Z_{E\to\hat 1}(1) \,.
\label{Zrel}
\end{equation}
Thus, if we understand the mixing of the chromo-electric operator in
(\ref{ZEdef}), we also understand the mixing of the kinetic energy
with the identity operator. As we shall see, calculating
$Z_{D^2\to\hat 1}$ by means of the relation (\ref{Zrel}) is much more
efficient than a direct calculation.

\begin{figure}
\epsfxsize=6cm
\centerline{\epsffile{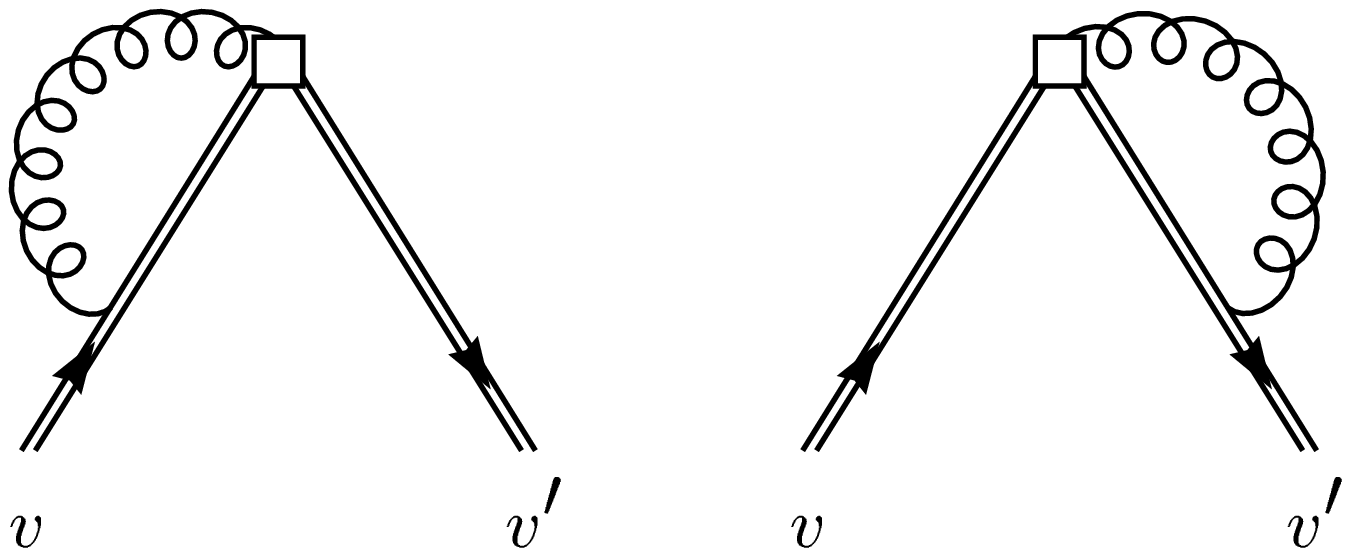}}
\caption{One-loop diagrams that could contribute to the mixing of the
chromo-electric operator with lower-dimensional operators. The
velocity-changing chromo-electric operator is represented by a white
square.}
\label{fig:1loop}
\end{figure}

Since the gluon field-strength is an antisymmetric Lorentz tensor,
the only possible dimension-three operator appearing on the
right-hand side of (\ref{ZEdef}) contains both velocities, $v$ and
$v'$. Consider now the diagrams shown in Fig.~\ref{fig:1loop}, which
could potentially give rise to operator mixing at the one-loop order.
Let us assume that we regulate the UV divergences of these diagrams
in a Lorentz-invariant way, for instance, by introducing a
Pauli--Villars cutoff. Since one leg of the gluon must be attached to
the chromo-electric operator, there is no way to obtain a dependence
on both heavy-quark velocities; the first diagram can only depend on
$v$, while the second one can only depend on $v'$. Hence, because of
Lorentz invariance the renormalization constant $Z_{E\to\hat 1}$ must
vanish at the one-loop order, and so does $Z_{D^2\to\hat 1}$. The
relation (\ref{Zrel}) implied by the virial theorem thus provides for
a very simple explanation of the apparently accidental cancellations
between the quadratic divergences of the one-loop diagrams in
Fig.~\ref{fig:kinop}, observed in \cite{MNS}.

\begin{figure}
\epsfxsize=10.5cm
\centerline{\epsffile{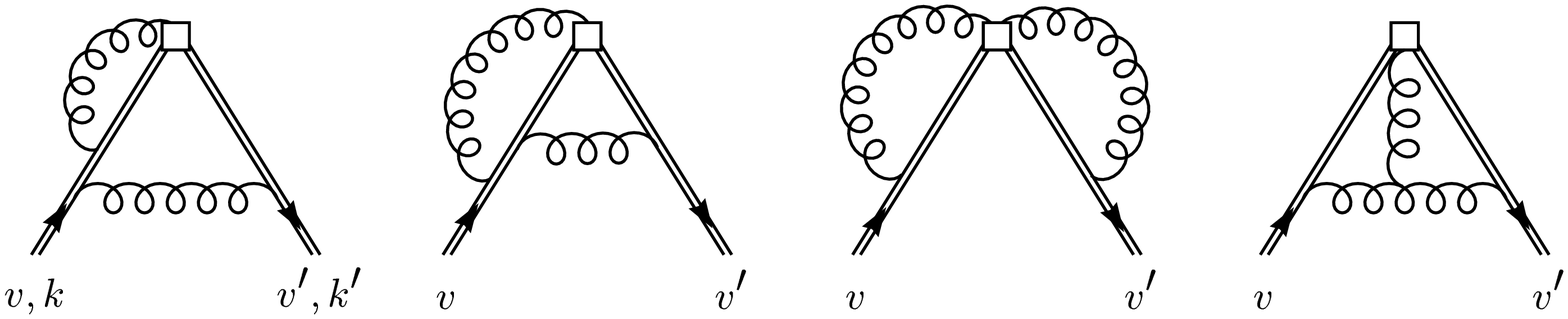}}
\caption{Two-loop diagrams that could contribute to the mixing of the
chromo-electric operator with lower-dimensional operators. Not shown
are two copies of the first two diagrams with the ``outer'' gluon
attached to the other heavy-quark line.}
\label{fig:2loop}
\end{figure}

On the other hand, the above argument is clearly linked to the
one-loop order. Let us, therefore, study the problem at the level of
two loops. Then there are four diagrams that could potentially lead
to a dependence on both heavy-quark velocities. They are shown in
Fig.~\ref{fig:2loop}. (Notice that this is a vast simplification over
the direct calculation of the two-loop mixing of the kinetic
operator.) We need to calculate the UV divergences of these diagrams
in a regularization scheme with a dimensionful cutoff. However, we
have to be careful not to spoil Lorentz or gauge invariance when
introducing the cutoff. For the case at hand, this can be done using
dispersion relations. Let us consider the matrix element of the
chromo-electric operator evaluated between quark states with residual
momenta $k$ and $k'$. (The nature of the external states is
irrelevant, since we are only interested in the UV behaviour of the
diagrams.) This matrix element has discontinuities if either $v\cdot
k$ or $v'\cdot k'$ are positive, since then physical intermediate
states can be excited. The Feynman amplitudes corresponding to the
diagrams in Fig.~\ref{fig:2loop} satisfy a dispersion representation
of the form\footnote{Once the double integral is regulated in the UV
region, there are no subtraction terms in the dispersion relation.}
\begin{equation}
   A = \int\mbox{d}\omega\int\mbox{d}\omega'\,
   \frac{\rho(\omega,\omega')}
   {(\omega-v\cdot k-i\epsilon)(\omega'-v'\cdot k'-i\epsilon)} \,,
\label{disp}
\end{equation}
where $\rho(\omega,\omega')$ is the double spectral density. The
variables $\omega$ and $\omega'$ have a concrete physical meaning as
the energies of the intermediate states, measured with respect to the
heavy-quark mass. A mixing with the identity operator would show up
as a quadratic divergence of the dispersion integral arising from the
region of large energies. Introducing an UV cutoff in the dispersion
integral regulates these divergences without spoiling any of the
symmetries of the theory; in particular, Lorentz and gauge invariance
are preserved.

Below, we shall investigate two different prescriptions how to
regulate the dispersion integral. The most direct way is to introduce
a hard UV cutoff, i.e.
\begin{equation}
   A^{\rm reg} = \int\limits^{\lambda}\!\mbox{d}\omega\!
   \int\limits^{\lambda}\!\mbox{d}\omega'\,
   \frac{\rho(\omega,\omega')}{\omega\omega'} \,.
\label{Areg}
\end{equation}
As we are interested in studying the UV behaviour of the operator
matrix element, we have set the external momenta to zero (there are
no IR divergences). For dimensional reasons, the result will then be
quadratic in the cutoff $\lambda$. Another scheme is to perform a
double Borel transformation of the original integral in the external
variables $v\cdot k$ and $v'\cdot k'$. This leads to
\begin{equation}
   A^{\rm Borel}
   = \frac{1}{M M'}\int\mbox{d}\omega\int\mbox{d}\omega'\,
   \rho(\omega,\omega')\,e^{-\omega/M-\omega'/M'} \,,
\end{equation}
corresponding to a ``soft'' exponential cutoff. It is most natural to
set the Borel parameters equal: $M=M'=\lambda$, in which case the
result will be proportional to $\lambda^2$.

To evaluate the two-loop diagrams in Fig.~\ref{fig:2loop}, we work in
the Feynman gauge; the sum of all diagrams is gauge independent.
Using the Feynman rules of the HQET, it is then straightforward to
see that the first two diagrams vanish. The third diagram is easily
calculated, since it factorizes into two one-loop integrals. Using
dimensional regularization, with $d$ being the number of space-time
dimensions, we obtain
\begin{eqnarray}
   &&- 2 C_F C_A\,\frac{g_s^4}{(4\pi)^d}\,\Gamma^2(3-d)\,
    \Gamma^2(d/2-1)\,(-2 v\cdot k)^{d-3} (-2 v'\cdot k')^{d-3}
    \nonumber\\
   &&\qquad\times (v^\mu v'^\nu - v^\nu v'^\mu)\,
    \bar u_{v'}(k')\,u_v(k) \,.
\end{eqnarray}
Here $C_F=\frac 12(N_c^2-1)/N_c$ and $C_A=N_c$ are the eigenvalues of
the quadratic Casimir operator in the fundamental and the adjoint
representations, and $u_v(k)$ is a heavy-quark spinor. The external
momenta $k$ and $k'$ act as IR regulators. The result is real if
$v\cdot k<0$ and $v'\cdot k'<0$. However, as expected there are
discontinuities for positive values of these variables. The
corresponding double spectral density is finite in four dimensions
and given by
\begin{equation}
   - C_F C_A\,\frac{\alpha_s^2}{2\pi^2}\,
   \theta(\omega)\,\theta(\omega')\,\omega\omega'\,
   (v^\mu v'^\nu - v^\nu v'^\mu)\,\bar u_{v'}(k')\,u_v(k) \,.
\end{equation}
The calculation of the last diagram is more involved, since this is a
genuine two-loop diagram with a non-trivial dependence on the
variable $v\cdot v'$. We shall first discuss the result obtained in
the scheme based on using a double Borel transformation. For this
case, efficient techniques to calculate two-loop integrals involving
two heavy quarks moving at different velocities have been developed
in \cite{twoloop}. In the case of equal Borel parameters
($M=M'=\lambda$) we find, after combining the results for the two
diagrams, that the renormalization factor defined in (\ref{ZEdef}) is
given by
\begin{equation}
   Z_{E\to\hat 1}^{\rm Borel}(v\cdot v') = -C_F C_A\,
   \frac{\alpha_s^2}{4\pi^2}\,F(v\cdot v')\,\lambda^2 \,,
\end{equation}
where
\begin{equation}
   F(w) = \frac{3w+2}{(w^2-1)^{3/2}}\,\Big[ L_2(-w_+) - L_2(-w_-)
   \Big] + \frac{2w+3}{w^2-1}\,\ln[2(w+1)] - \frac{1}{w+1} \,.
\end{equation}
Here $w_\pm=w\pm\sqrt{w^2-1}$, and $L_2(x)=-\int_0^x \frac{{\rm
d}y}{y}\ln(1-y)$ is the dilogarithm. Evaluating this expression in
the equal-velocity limit, using that $F(1)=\frac 13(1+2\ln 2)$, we
obtain from (\ref{Zrel}) the two-loop renormalization constant for
the mixing of the kinetic energy with the identity operator. The
result is
\begin{equation}
   Z_{D^2\to\hat 1}^{\rm Borel} = C_F C_A\,
   \frac{\alpha_s^2}{4\pi^2}\,(1+2\ln 2)\,\lambda^2 \,.
\label{ZBorel}
\end{equation}
The precise numerical coefficient of the quadratic divergence depends
on the way in which the UV cutoff $\lambda$ is introduced; below we
will derive the result for another scheme. However, the fact that we
have found a regularization scheme that preserves Lorentz and gauge
invariance, and in which the kinetic energy mixes with the identity
operator at the two-loop order, proves that such a mixing is present
in general. Beyond the one-loop order, it is not forbidden by any
symmetry.

Let us now elaborate on the result (\ref{ZBorel}). We have calculated
the generalization of this relation to the case of unequal Borel
parameters. Then
\begin{equation}
   Z_{D^2\to\hat 1}^{\rm Borel} = C_F C_A\,
   \frac{\alpha_s^2}{4\pi^2}\,M M'\,G(M/M') \,,
\label{ZMM}
\end{equation}
where
\begin{equation}
   G(x) = \frac{2(2-3x+2x^2)}{x}\,\ln(1+x)
   + \frac{2x^2(1-3x-2x^2)}{(1+x)^3}\,\ln x
   - \frac{1-6x+x^2}{(1+x)^2} \,.
\end{equation}
The function $G(x)$ is symmetric under the exchange $x\leftrightarrow
1/x$ and obeys $G(1)=1+2\ln 2$. Expression (\ref{ZMM}) is useful,
since it allows us to derive the double spectral density in the
dispersion integral (\ref{disp}). It is obtained by performing
another double Borel transformation, now in the variables $-1/M$ and
$-1/M'$, with Borel parameters $1/\omega$ and $1/\omega'$ \cite{Rad}.
The calculation is straightforward and leads to
\begin{equation}
   Z_{D^2\to\hat 1} = C_F C_A\,\frac{\alpha_s^2}{4\pi^2}
   \int\mbox{d}\omega\int\mbox{d}\omega'\,
   \frac{\bar\rho(\omega,\omega')}{\omega\omega'} \,,
\end{equation}
where
\begin{equation}
   \bar\rho(\omega,\omega') = \theta(\omega)\,\theta(\omega')\,
   \left\{ \omega\omega' + 5\omega_<^2
   + 2(3\omega\omega'-\omega^2-\omega'^2)\,\ln\bigg( 1
   - \frac{\omega_<}{\omega_>} \bigg) \right\} \,,
\end{equation}
with $\omega_<=\mbox{min}(\omega,\omega')$ and
$\omega_>=\mbox{max}(\omega,\omega')$. With the explicit result for
the double spectral density at hand, we can now evaluate, according
to (\ref{Areg}), the renormalization factor in the scheme with a hard
UV cutoff. The result is
\begin{equation}
   Z_{D^2\to\hat 1}^{\rm reg} = C_F C_A\,\frac{\alpha_s^2}{4\pi^2}
   \bigg( \frac{\pi^2}{3} - 1 \bigg)\,\lambda^2 \,.
\end{equation}
As expected, the numerical coefficient in front of the quadratic
divergence is different from the case of the Borel regulator in
(\ref{ZBorel}), but the general structure of the result is the same
in both regularization schemes.

We emphasize that in deriving that there is no mixing at the one-loop
order it is crucial that the UV cutoff be introduced in a fully
Lorentz-invariant way; in particular, for our arguments to hold it is
not sufficient to choose a regularization scheme that respects only
rotational invariance in the rest frame of the heavy hadron. This is
relevant when comparing our results to some previous calculations, in
which a non-vanishing contribution to $Z_{D^2\to\hat 1}$ was obtained
at $O(\alpha_s)$. In the case of the lattice regularization, the
one-loop mixing of the kinetic energy with the identity operator
arises because the lattice breaks Lorentz invariance (i.e., Euclidean
rotational invariance). The result for the renormalization factor
$Z_{D^2\to\hat 1}$ in this scheme is \cite{MMS}
\begin{equation}
   Z_{D^2\to\hat 1}^{\rm latt} = - k_{\rm latt}\,C_F\,
   \frac{\alpha_s}{\pi}\,\lambda^2 \,,
\label{Zlatt}
\end{equation}
where $\lambda=\pi/a$ is the UV cutoff ($a$ is the lattice spacing),
and the numerical coefficient $k_{\rm latt}$ is given by
\begin{equation}
   k_{\rm latt} = \frac{1}{2\pi^3} \int\limits_{-\pi}^\pi\!
   \mbox{d}^3k\,\frac{1}{\sqrt{(1+A)^2-1}} \,,\qquad
   A = \sum_{i=1}^3\,(1-\cos k_i) \,.
\end{equation}
The appearance of the quantity $A$ shows that the lattice
regularization breaks both Lorentz and rotational invariance. Things
are slightly more intricate with the scheme adopted in \cite{Bigsr},
however. The definition for the parameter $\lambda_1^H(\lambda)$
suggested there is equivalent to introducing a cutoff on the spatial
momentum ($|{\bf k}|<\lambda$) in the rest frame of the hadron
containing the heavy quark.\footnote{The author is grateful to
M.~Shifman for pointing out this fact, and for suggesting the
derivation of (\protect\ref{ZBigi}) along the lines discussed here.}
Although working in the rest frame is quite natural when dealing with
heavy quarks, this prescription breaks Lorentz invariance in that it
specifies a particular reference frame, thereby introducing a
dependence on the velocity $v$ through the regularization procedure.
Then the arguments that led us to conclude that the one-loop diagrams
in Fig.~\ref{fig:1loop} have no quadratic divergences are no longer
valid. The second diagram, which explicitly depends on $v'$, now
acquires an implicit dependence on $v$ through the regularization,
and thus it can contribute to $Z_{E\to\hat 1}$ in (\ref{ZEdef}).
After a straightforward calculation, we find that
\begin{equation}
   Z_{E\to\hat 1}(v\cdot v') = C_F\,
   \frac{\alpha_s}{2\pi}\,H(v\cdot v')\,\lambda^2 \,,
\end{equation}
where
\begin{equation}
   H(w) = \frac{w-r(w)}{w^2-1} \,,\qquad
   r(w) = \frac{1}{\sqrt{w^2-1}}\,\ln\left(w+\sqrt{w^2-1}\right) \,.
\end{equation}
The same result is obtained if the cutoff is introduced in the rest
frame of the final-state hadron. Using that $H(1)=\frac 32$, we
obtain from (\ref{Zrel})
\begin{equation}
   Z_{D^2\to\hat 1} = - C_F\,\frac{\alpha_s}{\pi}\,\lambda^2 \,,
\label{ZBigi}
\end{equation}
which agrees with the result derived in \cite{Bigsr}. The same result
can be recovered from (\ref{Zlatt}) by taking the ``continuum limit''
$A\to\frac 12\,{\bf k}^2$ and imposing the cutoff on $|{\bf k}|$
(i.e.\ by integrating over a sphere rather than a cube), in which
case rotational invariance is restored, and $k_{\rm latt}\to 1$.
Since the lattice version of the HQET is formulated for heavy quarks
at rest, Lorentz invariance is broken even in the ``continuum
limit''.

\section{Conclusions}

Using the virial theorem of the heavy-quark effective theory, we have
shown that the mixing of the operator for the heavy-quark kinetic
energy with the identity operator is forbidden at the one-loop order
by Lorentz invariance. This resolves the puzzle of the ``invisible
renormalon'' raised in \cite{MNS}, which consisted in the observation
of an apparently accidental vanishing of the mixing at the one-loop
order in several Lorentz-invariant cutoff regularization schemes, and
the associated absence of an UV renormalon singularity in the bubble
approximation. However, the mixing is not protected in general by any
symmetry, and we have shown explicitly that it does occur at the
two-loop order. This means that the parameter $\lambda_1^H$ of the
heavy-quark effective theory is not directly a physical quantity, but
requires a non-perturbative subtraction.

Our results for the two-loop renormalization of the kinetic energy
imply that, in any regularization scheme with a dimensionful cutoff
parameter $\lambda$ that preserves Lorentz and gauge invariance,
there is an additive, quadratically divergent contribution to the
parameter $\lambda_1^H(\lambda)$ of the HQET, such that
\begin{equation}
   \frac{\mbox{d}}{\mbox{d}\lambda^2}\,\lambda_1^H(\lambda)
   = \frac{\mbox{d}}{\mbox{d}\lambda^2}\,Z_{D^2\to\hat 1}
   = k\,C_F C_A\,\frac{\alpha_s^2(\lambda)}{4\pi^2}
   + O(\alpha_s^3) \,.
\end{equation}
For the two regularization schemes considered here, the numerical
coefficient $k$ takes the values $(1+2\ln 2)\approx 2.39$ and $(\frac
13\pi^2-1)\approx 2.29$. Note that this renormalization-group
equation is different from the one derived in \cite{Bigsr}, which
contains a one-loop contribution of the form
$-C_F\,\alpha_s(\lambda)/\pi$ on the right-hand side, in accordance
with (\ref{ZBigi}). The reason is that the ``physical'' definition of
the parameter $\mu_{\pi,H}^2(\lambda)=-\lambda_1^H(\lambda)$
suggested by these authors is equivalent to introducing a cutoff on
the spatial momentum in the rest frame of the heavy hadron and thus
breaks Lorentz invariance. (Although it is perfectly consistent, this
definition may therefore not be the most convenient one.) The values
of the kinetic energy defined in different regularization schemes can
be related to each other using perturbation theory. From our results,
it follows that at the one-loop order there is a universal relation
between the parameter $\mu_{\pi,H}^2(\lambda)$ defined in
\cite{Bigsr} and $\lambda_1^H(\lambda)$ defined in any
Lorentz-invariant regularization scheme, which reads
\begin{equation}
   \mu_{\pi,H}^2(\lambda) = -\lambda_1^H(\lambda)
   + C_F\,\frac{\alpha_s(\lambda)}{\pi}\,\lambda^2
   + O(\alpha_s^2) \,.
\label{mulamrel}
\end{equation}
This implies that the lower bound for the kinetic energy of a heavy
quark inside a meson, $\mu_{\pi,{\rm
mes}}^2(\lambda)>0.36~\mbox{GeV}^2$, which by construction holds for
the definition proposed in \cite{Bigsr}, becomes weaker when the
kinetic energy is regulated in a Lorentz-invariant way. Even if the
cutoff $\lambda$ is chosen as low as 1~GeV, the extra term in
(\ref{mulamrel}) amounts to about 0.15~GeV$^2$, implying that the
lower bound for $-\lambda_1^{\rm mes}(\lambda)$ is at most
0.2~GeV$^2$. This observation is in accordance with arguments
presented in \cite{KLWG}, where a similar conclusion has been
reached.

As mentioned in the introduction, the presence of a quadratic
divergence entails that the parameter $\lambda_1^H$ of the HQET
requires a non-perturbative subtraction to be defined and as such is
not directly a physical parameter. Since in Lorentz-invariant
regularization schemes the operator mixing occurs only at the
next-to-leading order, one may hope that in practice the ambiguities
related to the choice of the subtraction scheme will be small.
Nevertheless, one has to be careful when comparing the values of
$\lambda_1^H$ obtained using different theoretical methods. For
instance, QCD sum-rule determinations of $\lambda_1^H$ for the
ground-state heavy mesons \cite{ElSh}--\cite{MN96} and baryons
\cite{Cola} are performed in a scheme based on dimensional
regularization, i.e.\ without a dimensionful UV cutoff. Strictly
speaking, the results are then affected by a renormalon ambiguity
problem, i.e.\ there is an intrinsic arbitrariness of order
$\Lambda_{\rm QCD}^2$, which would show up when higher-order
perturbative corrections to the sum rules would be calculated beyond
the bubble approximation. Still, these results can consistently be
used in connection with Wilson coefficients calculated to the same
accuracy. Therefore, the sum-rule calculations may be compared with
the results of phenomenological determinations of $\lambda_1^{\rm
mes}$ from a combined analysis of inclusive decay rates and moments
of the decay spectra in decays of beauty and charm mesons
\cite{GKLW}--\cite{Cher}. On the other hand, when $\lambda_1^H$ is
determined using lattice gauge theory, the presence of a dimensionful
regulator (the lattice spacing) requires a non-perturbative
subtraction, and the results depend on how this subtraction is
performed. A ``physical prescription'' for such a subtraction has
been suggested in \cite{GiMS}.

\vspace{0.3cm}
{\it Acknowledgement:\/}
It is a pleasure to thank Martin Beneke, Chris Sachrajda and Misha
Shifman for many useful discussions and suggestions.

\end{document}